\documentstyle[12pt]{article}
\newcounter{mycount}

\newcommand{\bee}{\begin{eqnarray}}
\newcommand{\eee}{\end{eqnarray}}
\newcommand{\be}{\begin{eqnarray}}
\newcommand{\ee}{\end{eqnarray}}
\newcommand\nn{\nonumber \\}
\newcommand\half{\frac{1}{2}}

\newcommand\hm{\hat{m}}
\newcommand\hn{\hat{n}}
\newcommand\as{
{\fr s}}
\newcommand\aA{
{\fr a}}
\newcommand\aB{
{\fr b}}
\newcommand\aD{
{\fr c}}
\newcommand\at{
{\fr t}}
\newcommand\un{{
{n}}}

\font\frtnfr=eufm10   scaled\magstep1
\font\twlfr=eufm10
\font\tenfr=eufm10

\newfam\frfam
\textfont\frfam=\frtnfr
\scriptfont\frfam=\twlfr
\scriptscriptfont\frfam=\tenfr
\def\fr{\fam\frfam}

\font\frtnopen=msbm10  scaled\magstep2
\font\twlopen=msbm10
\font\tenopen=msbm10

\newfam\openfam
\textfont\openfam=\frtnopen
\scriptfont\openfam=\twlopen
\scriptscriptfont\openfam=\tenopen

\font\frtnsf = cmss12 scaled\magstep1
\font\twlsf = cmss10
\font\tensf = cmss9

\newfam\Scfam
\textfont\Scfam = \frtnsf
\scriptfont\Scfam = \twlsf
\scriptscriptfont\Scfam = \tensf

\begin{document}
\renewcommand{\theequation}{\arabic{equation}}
\bibliographystyle{nphys}
\setcounter{equation}{0}

\title{
\begin{flushright}
{\small
hep-th/0010239\\
FIAN/TD/16--00}\\
\end{flushright}
\vspace{5mm}
Conformal Higher Spin Currents in Any Dimension
and AdS/CFT Correspondence}

\author{S.E.Konstein, M.A.Vasiliev and V.N.Zaikin \\
  \\
               {\small \phantom{uuu}}
  \\
           {\it {\small} I.E.Tamm Department of Theoretical Physics,}
  \\
               {\it {\small} P. N. Lebedev Physical Institute,}
  \\
         {\it {\small} Leninsky Prospect 53, Moscow 117924, Russia.}
}

\date{}

\maketitle

\begin{abstract}
{The full list of conserved conformal higher
spin currents built from massless scalar and spinor fields is presented.
It is shown that, analogously to the relationship
between usual conformal and $AdS$ symmetries, the set of the
conformal higher spin
symmetry parameters associated with the conformal conserved currents
in $d$ dimensions is in the one-to-one correspondence with the result of the
dimensional reduction of the usual (i.e., non-conformal) higher
spin symmetry parameters in $d+1$ dimensions.}
\end{abstract}

\renewcommand{\theequation}{\arabic{equation}}
\setcounter{equation}{0}

\section {Introduction}

Usual lower spin conserved  currents
admit a natural extension to the higher spin case in any space-time
dimension. Historically first examples
of the higher spin currents
were known as zilch currents \cite{zilch}. More examples were given in
\cite{currents,ex}.
The full list of conserved higher spin currents, including those
containing explicit dependence on the spacetime coordinates, was given
recently in \cite{MV} where it was shown in particular that various
types of conserved currents  $J^\un{}_{;a(\at),b(\as)}$
associated with the integer spin $s=\as+1$ are vector fields (index $\un$)
taking values in the representations of the Lorentz group
$SO(d-1,1)$ described by the traceless two-row Young diagrams

\vskip -5mm
\bigskip
\centerline{\small $\as$}
\vskip -10mm
\be
\label{dia}
\begin{picture}(20,50)
\put(33,35){\circle*{2}}
\put(25,35){\circle*{2}}
\put(17,35){\circle*{2}}
\put(25,25){\circle*{2}}
\put(17,25){\circle*{2}}
\put(33,25){\circle*{2}}
\put(00,40){\line(1,0){70}}
\put(00,30){\line(1,0){70}}
\put(50,30){\line(0,1){10}}
\put(60,30){\line(0,1){10}}
\put(70,30){\line(0,1){10}}
\put(00,20){\line(1,0){60}}
\put(00,20){\line(0,1){20}}
\put(10,20){\line(0,1){20}}
\put(40,20){\line(0,1){20}}
\put(60,20){\line(0,1){20}}
\put(50,20){\line(0,1){20}}
\end{picture}
\ee
\vskip -10mm
\centerline{$\at$}

\vskip 4mm
\noindent
with $0\le \at \le \as$. This means that the tensor
$
J^\un{}_{;a(\at ),b(\as)}$
is symmetric in the indices $b$ and $a$
separately\footnote{Following to \cite{V2},
we use  notation $a(\at )$
for  multiindices $a_1 a_2\,...\, a_\at$ subject to symmetrization
(by symmetrization we
mean the appropriate projector, i.e. symmetrization of a symmetric
tensor leaves it unchanged).
Multiindex $a(\at )a(\as )$ is equivalent to $a(\at+\as)$.
We use notation
$A^{a(\at )}B_{a(\at )}$ for $A^{a_1 \ldots a_\at} B_{a_1 \ldots a_\at}$ with
symmetrized tensors $A$ and $B$. The
tensor indices are raised and lowered by the flat Minkowski metric
$\eta_{ab}$.}
($a,b,n = 0,\,...\,,\, d-1 $, where $d$ is the space-time  dimension),
obeys the antisymmetry property
\be
\label{asym}
J^\un{}_{;a(\at -1)b(1), \, b(\as)}=0
\ee
and has zero contractions of indices $a$ and/or $b$.
It is enough to require
\be
\label{tr}
J^\un{}_{;a(\at ),b(\as -2)c}{}^c=0
\ee
since from (\ref{asym}) it follows then that any other contraction
of indices $a$ and/or $b$ is zero.

It is convenient to consider the currents
\be
J^\un ({\xi}) =
\!J^{\un}{}_{;}{}^{a(\at )}{}_{,}{}^{b(\as)}
\xi_{a(\at ),b(\as)}\,,\quad\!
\ee
where $\xi_{a(\at ),b(\as)}$ are some constant parameters which transform
under the irreducible representation of Lorentz group
corresponding to the traceless two-row Young diagram. The role of the
parameter $\xi_{a(\at ),b(\as)}$ is that it projects the current to the
appropriate irreducible representation of the Lorentz group.

The current $J^\un ({\xi})$ conserves if
\be
\partial_\un J^\un ({\xi})=0\,.
\ee

In the cases discussed below,
a conserved current corresponding to the one-row Young diagram
admits a representation
\be
\label{JT}
J^{n}{}_{ ;\,\,,}{}^{  a(\as)}
\xi_{a(\as)}= T^{n\,a(\as)} \xi_{a(\as)},
\ee
where $T^{a(\as +1 )}$ is a totally symmetric conserved
current
\be
\partial_n T^{n a(\as)}=0\,.
\ee
Given conserved
current $T^{n(\as +1 )}$, it  allows one to construct the set of
integer spin conserved currents
\cite{MV},
\be
\label{JTB}
\!J^\un ({\xi}) =
T^{\un l(\as)}x^{m(\at )}
\xi_{m(\at ),l(\as)},\quad\!
\ee
where we use the shorthand notation
\be
\label{xs}
x^{m(\as )} =
x^{m_1} \ldots x^{m_\as} \,.
\ee
These generalize the usual spin one current $T^n$, stress tensor
$T^{nm}$ and angular momentum current
\be\label{lor}
M^{\un}{\,}^{ b,a}=T^{\un a} x^b - T^{\un b} x^a\,
\ee
to an arbitrary integer spin. A form of the symmetric current
$T^{a(s)}$ depends on a particular matter system. For scalar and
spinor fields some examples are given in \cite{ex,MV}
for any spin. Note that the
current $T^{a(s)}$ is not required to be traceless in this construction.
However, the terms with double traces in $T^{a(s)}$ do not contribute to
(\ref{JTB}) because  $\xi_{a(\at ),b(\as)}$ is traceless. The situation with
supercurrents is analogous \cite{MV}.

It is well known that if the stress tensor is traceless, this
implies
a larger conformal symmetry. Indeed, in that case the currents
associated with the dilatation (scale transformation)
\be\label{del}
T_1^\un = T^{\un k} x_k
\ee
and  special conformal transformations
\be\label{sct}
K_1^{\un a} = T^{\un a} x^2 -2 T_1^\un x^a   \qquad (x^2 = \eta_{ab}x^a x^b )
\ee
conserve.
The currents (\ref{lor}), (\ref{del}) and (\ref{sct})
and $T^{\un a}$ itself exhaust all those currents that
can be constructed from the traceless stress tensor $T^{\un a}$.

It was conjectured in \cite{MV} that
a similar phenomenon takes place for all spins,
implying larger conformal higher spin symmetries for the case
of traceless  tensors (\ref{JT}).  The aim of this paper is to
give the full list of conserved conformal higher spin currents
that can be built from scalar and spinor fields.
To elucidate a general pattern we first consider the simplest
nontrivial example of the spin three current.

\section{Spin Three Example}

Let the symmetric tensor
$T^{abc}$ be traceless and conserved
\be
\partial_a T^{abc} =0\,, \qquad T^{ab}{}_b =0\,.
\ee
First, one observes that the traceless tensor
\be
T_1^{\un \,\,a} &=& T^{\un a k} x_k
\ee
 conserves. It allows one to construct
conserved currents analogous to
(\ref{lor}), (\ref{del}) and (\ref{sct}):
\be
T_2^{\un } &=& T_1^{\un \,\,k} x_k
\ee
\be
M_{1}^{\un}{}_{\,\,}{}^{b, a}&=& T_1^{\un \,\,a}x^b- T_1^{\un \,\,b }x^a,
\ee
and
\be
K_{1}^{\un \,\,a} = T_1^{\un \,\,a} x^2 -2 T_2^{\un} x^a\,,\qquad
x^2 \equiv x_bx^b\,.
\ee

In addition, one can check that the following  currents conserve
\be
K_{2}^{\un \,\,a_1 a_2}&=& T^{\un a_1 a_2} (x^2)^2
      -2 \left(
               T_1^{\un \,\,a_1} x^{a_2}
             + T_1^{\un \,\,a_2} x^{a_1} \right) x^2
     +4 T_2^\un x^{a_1} x^{a_2}            \,,
\ee
\be
K_{1}^{\un \,\,a_1 a_2}&=& T^{\un a_1 a_2} x^2
                       -T_1^{\un \,\,a_1} x^{a_2}
                       -T_1^{\un \,\,a_2} x^{a_1}
                       +\frac 2 d T_2^\un  \eta^{a_1 a_2},
\ee
\be
\label{lc1}
M^{\un}{}_{\,\,}{}^{b, a_1 a_2}&=&
                               2 T^{\un a_1 a_2} x^{b    }
                               - T^{\un a_1 b  } x^{a_2  }
                               - T^{\un b a_2  } x^{a_1  }          \nn
    &+& \frac 1 {d-1} \left(
                         2 \eta^{a_1 a_2} T_1^{\un \,\,b   }
                         - \eta^{a_1 b  } T_1^{\un \,\,a_2 }
                         - \eta^{b a_2  } T_1^{\un \,\,a_1 } \right),
\ee
\be
M_{1}^{\un}{}_{\,\,}{}^{b, a_1 a_2}&=&
                     M^{\un}{}_{\,\,}{}^{b, a_1 a_2} x^2        \nn
 &-& 2 \left(
          M_{1}^{\un}{}_{\,\,}{}^{b, a_2} x^{a_1}
        + M_{1}^{\un}{}_{\,\,}{}^{b, a_1} x^{a_2} \right)           \nn
   &-& \frac 2 {d-1} \left(
                         2 \eta^{a_1 a_2} M_{1}^{\un}{}_{\,\,}{}^{k, b  }x_k
                         - \eta^{a_1 b  } M_{1}^{\un}{}_{\,\,}{}^{k, a_2}x_k
                         - \eta^{b a_2  } M_{1}^{\un}{}_{\,\,}{}^{k, a_1}x_k
                   \right)
\ee
and
\be
\label{lc2}
M^{\un}{}_{\,\,}{}^{b_1 b_2, a_1 a_2} &=&
                                           2 T^{\un a_1 a_2}x^{b_1}x^{b_2}
                                          - T^{\un b_1 a_2}x^{a_1}x^{b_2}
                                          - T^{\un a_1 b_2}x^{b_1}x^{a_2} \nn
                                         &+& 2 T^{\un b_1 b_2}x^{a_1}x^{a_2}
                                          - T^{\un a_2 b_2}x^{a_1}x^{b_1}
                                          - T^{\un a_1 b_1}x^{a_2}x^{b_2} \nn
&-& \frac 1 {d-1} \left (
                   2  \eta^{a_1 a_2} K_{1}^{\un \,\,b_1 b_2}
                   -  \eta^{b_1 a_2} K_{1}^{\un \,\,a_1 b_2}
                   -  \eta^{a_1 b_2} K_{1}^{\un \,\,b_1 a_2}  \right. \nn
                  &+& \left. 2  \eta^{b_1 b_2} K_{1}^{\un \,\,a_1 a_2}
                   -  \eta^{a_2 b_2} K_{1}^{\un \,\,a_1 b_1}
                   -  \eta^{a_1 b_1} K_{1}^{\un \,\,a_2 b_2} \right)  \nn
&+& \frac 2 {d(d+1)} \left(
                   2 \eta^{a_1 a_2}  \eta^{b_1 b_2}
                   - \eta^{b_1 a_2}  \eta^{a_1 b_2}
                   - \eta^{a_1 b_1}  \eta^{b_2 a_2} \right) T_2^{\un }\,.
\ee
These currents have the following symmetry properties
\be
K_{2}^{\un \,\,a_1 a_2}=K_{2}^{\un \,\,a_2 a_1},\qquad
K_{1}^{\un \,\,a_1 a_2}=K_{1}^{\un \,\,a_2 a_1},
\ee
\be
   M^{\un}{}_{\,\,}{}^{b, a_1 a_2} = M^{\un}{}_{\,\,}{}^{b, a_2 a_1}=
- M^{\un}{}_{\,\,}{}^{a_1, a_2 b}
- M^{\un}{}_{\,\,}{}^{a_2, a_1 b}\,,
\ee
\be
   M_1^{\un}{}_{\,\,}{}^{b, a_1 a_2} = M_1^{\un}{}_{\,\,}{}^{b, a_2 a_1}=
- M_1^{\un}{}_{\,\,}{}^{a_1, a_2 b}
- M_1^{\un}{}_{\,\,}{}^{a_2, a_1 b}\,,\qquad
\ee
\be
M^{\un}{}_{\,\,}{}^{b_1 b_2, a_1 a_2} =
M^{\un}{}_{\,\,}{}^{b_1 b_2, a_2 a_1} =
M^{\un}{}_{\,\,}{}^{b_2 b_1, a_1 a_2}\,,\qquad \nn
M^{\un}{}_{\,\,}{}^{b_1 b_2, a_1 a_2} +
M^{\un}{}_{\,\,}{}^{b_1 a_1, a_2 b_2} +
M^{\un}{}_{\,\,}{}^{b_1 a_2, b_2 a_1} =0.
\ee

Note that only the original current $T^{abc}$ and the Lorentz-type
currents (\ref{lc1}) and (\ref{lc2}) conserve in the nonconformal case
considered in \cite{MV}
(after adding appropriate terms containing the traces $T^{abc}\eta_{bc}$
that are not supposed to be zero in the non-conformal case).
All other currents in the above list
require the ``stress tensor" $T^{abc}$ to be traceless and generalize
the dilatation and special conformal currents of the usual conformal case.
{}From the analysis of this section it is seen that different currents
classify according to their (i) symmetry properties,
(ii) a degree of homogeneity in $x^a$ and
(iii) a highest power of $x^2$.

\section{ Conformal Currents of Any Spin}
\label{Higher Spin Conformal Currents}

Let us first present the full list of conserved higher spin currents
for an arbitrary integer spin $s=\as+1$.
Let $T^{ a(s)}$ be totally symmetric and traceless, i.e.
\be\label{2}
T^{bc a(s -2)}\eta_{bc}=0.
\ee

Provided that $T^{ a(s)}$ is conserved,
\be\label{1}
\frac \partial {\partial x^n} T^{n \, a(\as)} =0\,,
\ee
 one can construct a three-parametric  family of the
conserved tensors.

First one constructs a family of lower rank conserved
totally symmetric traceless tensors
\be\label{29}
T_{q}^{a(s-q)}\stackrel {def} =
T^{a(s-q)b(q)} x_{b(q)}\,.
\ee
Obviously, these tensors are symmetric
and themselves satisfy the conditions
(\ref{1}),  (\ref{2}). Therefore one can use the tensors
$T_{q}^{a(s-q)}$ the same way as the original spin $s-q$
tensors $T^{a(s-q)}$.

The analysis of the spin 2 and spin 3 cases indicates
another source of multiplicity of currents  due to the possibility to
multiply by powers of $x^2$. Taking into account that
the conservation condition is homogeneous in $x^a$ one looks for
a current of the form
\be
J^\un  = \sum_{\aD=0}^\aA (x^2 )^{\aA-\aD} C(\aA,\aD,\as,\at)
T_\aD^{\un \,a(\as-\aD)} \xi_{b(\at)\,, a(\as-\aD)b(\aD)} x^{b(\at+\aD)}
\ee
with some coefficients $C(\aA,\aD,\as,\at)$ and arbitrary parameters
$\xi_{b(\at)\,, a(\as)}$ corresponding to the traceless
two-row Young diagram. An elementary
computation shows that the current $J^\un $ conserves provided that
\be
C(\aA,\aD,\as,\at ) =
(-2)^\aD \frac{(\as-\aD-\at)!}{\aD ! (\aA-\aD)! (\as-\aD)!}
\ee
and $\aA + \at \leq \as $.

Taking into account that analogous formula for currents $T_\aB$ (\ref{29})
also leads to the conserved currents one arrives at the final result

\be
\label{Jxi}
J^\un_{\as,\at,\aA,\aB} (T|\xi ) =
\sum_{\aD=0}^\aA
(-2)^\aD \frac{(\as-\aB-\aD-\at)!}
{\aD ! (\aA-\aD)! (\as-\aB-\aD )!} (x^2 )^{\aA-\aD}
T_{\aB+\aD}^{\un \,a(\as-\aB-\aD)} \xi^\aA_{b(\at)\,,
a(\as-\aB-\aD)b(\aD)} x^{b(\at+\aD)}
\ee
for
\be
\label{ineq}
\aA+\aB+\at\leq \as\,.
\ee
Therefore, for any four non-negative integers $\aA$,$\aB$,$\as$ and $\at$
 satisfying the inequality
(\ref{ineq}) there is an independent conserved
current associated with the generalized stress tensor $T^{a(s)}$ and a
two-row Young diagram parameter
$\xi^\aA_{b(\at)\,, a(\as-\aB)}$. The label $\aA$ is introduced here
to distinguish between parameters that belong to the same representation
of the Lorentz group but are associated with different conserved currents.

The situation with half-integer higher spin supercurrents is
analogous. Let $R^{a(\as +1)}{}_\alpha$ be a totally symmetric
conserved tensor-spinor of spin $s=\as+3/2$
($\alpha$ is the spinor index in $d$ dimensions)
satisfying the  $\gamma$ transversality condition , i.e.
\be
\label{gtr}
\partial_a R^{a\,b(\as)}{}_\beta =0\,,\qquad
\left(\gamma_a\right)_\alpha{}^\beta R^{a \,b(\as)}{}_\beta  =0
\ee
with the $d$-dimensional Dirac $\gamma$ matrices
\be
\{\gamma_a ,\gamma_b\} =2 \eta_{ab}\,.
\ee
Since the tensor-spinor $R^{a(\as +1)}{}_\alpha $ is traceless
as a consequence of (\ref{gtr}), one can
apply the bosonic construction (\ref{Jxi}) to build the conserved
supercurrents
\bee
\label{Jxi1}
F^\un_{\as,\at,\aA,\aB} (R|\xi ) =
\sum_{\aD=0}^\aA&{}&
(-2)^\aD \frac{(\as -\aB-\aD-\at)!}{\aD! (\aA-\aD)!
(\as -\aB-\aD )!}\nonumber\\
&\times& (x^2 )^{\aA-\aD}
x^{b(\at +\aD)}
\xi^\aB_{b(\at )\,,
a(\as-\aB-\aD)b(\aD)}{}^\beta
R_{\aB+\aD}^{\un \,a(\as-\aB-\aD)}{}_\beta \,,
\eee
where
\be
R_{q}^{a(s -q)}\stackrel {def} = R^{a(s -q)b(q)} x_{b(q)}
\ee
and the tensor-spinor parameter is irreducible
\be
\xi^\aB_{b(u-1)a(1)\,, a(v)} {}^\beta =0\,,\qquad
         \xi^\aB_{b(u)\,, a(v)}{}^\beta (\gamma^a)_\beta{}^\alpha =0 \,.
\ee
Due to the $\gamma$-transversality condition (\ref{gtr}) the
multiplication of the right hand side of (\ref{Jxi1}) by the factor of
\be
\hat{x} = x^a \gamma_a
\ee
does not destroy the conservation property.
This gives another set of conformal higher spin supercurrents
\bee
\label{Jxi2}
G^\un_{\as,\at,\aA,\aB} (R|\xi ) =
\sum_{\aD=0}^\aA&{}&
(-2)^\aD \frac{(\as -\aB-\aD-\at)!}
{\aD! (\aA-\aD)! (\as -\aB-\aD )!}  \nonumber\\
&\times& (x^2 )^{\aA-\aD}
x^{b(\at +\aD)}
\Big (\xi^\aB_{b(\at )\,,
a(\as-\aB-\aD)b(\aD)} \hat{x}
R_{\aB+\aD}^{\un \,a(\as-\aB-\aD)}\Big )\,.
\ee
Note that the supercurrents $F$ and $G$
generalize the usual $Q$ and $S$ conformal supersymmetries.

The results of this
paper rely on the fact of the existence of the traceless
totally symmetric conserved tensors $T$
or $\gamma$-transversal supercurrents $R$.
The conserved traceless spin $s$ tensors built
in \cite{ex} from massless scalar and
spinor matter fields have the form
\bee
\label{bb}
T^{n(s )}&=&
\sum_{k=0}^{s }(-1)^{k}
\frac{(\frac{d}{2}-2)!(s+\frac{d}{2}-2)!}
{k!(s-k)!(\frac d 2 + k -2)!(s +\frac d 2 -k-2)!}\nn
&\times&
\Big[ \partial ^{n(k)}\varphi
{}~\partial^{n(s -k)}\varphi
-\frac{k(s -k)}{d+2s-4}
{}~\eta ^{n(2)}
{}~\partial ^{n(k-1)}~\partial _m\varphi ~\partial ^{n(s -k-1)}
{}~\partial ^m\varphi
\Big] \ldots \,
\eee
and
\bee
\label{ff}
{\cal J}^{n(s )}=
\sum_{k=0}^{s -1}
{}~\frac
{
(-1)^{k}
(\frac{d}{2}-1)!
(s+\frac{d}{2}-2)!
}
{
k!(s-k-1)!
(k+\frac{d}{2}-1)!
(s+\frac{d}{2}-k-2)!
}
\Big[ \partial ^{n(k)}\bar{\psi}
{}~\gamma^{n} ~\partial ^{n(s -k-1)}\psi
\nn
-~\frac{k(s -k-1)}{d+2s-4}
{}~\eta^{n(2)}\partial _{m}~\partial^{n(k-1)}\bar{\psi}
{}~\gamma^{n}
{}~\partial ^{m}
{}~\partial ^{n(s -k-2)}\psi \Big]\ldots \,,
\ee
where
\be
\partial^{a(p)} =\partial^{a_1} \ldots \partial^{a_p},
\ee
and dots stand for
discarded terms containing more than one factor of $\eta^{n(2)}$
(such terms do not contribute to (\ref{Jxi}) because
the parameter $\xi_{b(\at ),a(\as)}$ is traceless).

The half-integer spin $s$ gamma-traceless conserved tensor-spinor
has the form
\bee
\label{bf}
R^{a(s -1/2)}  &=& \sum_{l,n,p =0}^\infty (-1)^l
\frac{(\frac{d}{2} +p +l +n -1 )!}{l!n!p!2^p
(\frac{d}{2} +p +l -1 )! (\frac{d}{2} +p +n -1 )!}\nn
&{}&\Big [ (\frac{d}{2} +p +n -1 )
\delta (s -1/2-l-n-2p)
\eta^{a(2p)}
\partial^{a(l)}\partial^{b(p)}
\psi \partial^{a(n)}\partial_{b(p)} \varphi \nn
&-&\half\delta (s -1/2 -l -n -2p -1 )
\eta^{a(2p)}\gamma^{a(1)}
\partial^{a(l)}\partial^{b(p)} \gamma^{b(1)}
\psi \partial^{a(n)}\partial_{b(p+1)} \varphi \Big ]\,,
\eee
where, for odd $d$, the factorials have to be understood as
the corresponding $\Gamma$ functions,
$\delta (0) = 1$ and $\delta (n) = 0 $ for $n\neq 0$,
and we use the shorthand notation
\be
\label{eta}
\eta^{a(2p)} =\underbrace{\eta^{a(2)} \ldots \eta^{a(2)}}_p\,.
\ee
Note that the sum in (\ref{bf}) contains a finite number of
terms due to the factors of $\delta$.
The proof of the facts that
the supecurrent (\ref{bf}) is $\gamma-$transversal and conserves as
a consequence of the field equations for
the massless spinor and scalar
\be
\Box \varphi=0 \,,\qquad
\hat{\partial} \psi=0\,,\qquad\qquad
\Box\equiv\partial_b\partial^b\,,\qquad \hat{\partial}\equiv\partial_b\gamma^b
\ee
is straightforward but a little tedious. To the best of our knowledge
 the constructed
conformal higher spin supercurrents are new. We therefore
reproduce here their complete form (i.e., keeping all higher traces).

Let us note that the form of the irreducible currents (\ref{bb}),
(\ref{ff}) and (\ref{bf}) and, therefore, (\ref{Jxi}), (\ref{Jxi1})
and (\ref{Jxi2}) is fixed unambiguously by the conservation
and irreducibility conditions provided that there is no explicit
dependence on $x^m$ and an order of derivatives
is fixed to be minimal for a given rank
of the current. (This is not the case for the usual (non-conformal)
currents that are not required to be traceless or $\gamma$-transversal.)
Relaxing the latter condition one constructs
more conserved currents. For example, the following currents are obviously
conserved and irreducible:
\be
\Box T^{n(s)} \,,\qquad x^m\partial_m
T^{n(s)}  \,,\qquad \hat{ \partial} R^{a(s -1/2)}\,.
\ee
A less trivial example is
\bee
T^{n(s +1)} &=& \partial^{n(1)} T^{n(s)}
+\frac{1}{s+d -1} x^m\partial_m \partial^{n(1)} T^{n(s)}\nonumber\\
&+&\frac{s}{(s+d-1)(d+2s-3)}\Big (\eta^{n(2)}\Box - \partial^{n(2)}\Big )
\Big( T^{n(s-1)m}x_m \Big)\,.
\eee
However, we conjecture that all the
conserved successors containing derivatives of the
original currents do not give
rise to new charges, thus describing various improvements of the original
currents.

Indeed, the conserved currents give rise to the conserved charges
in the standard way by virtue of the integration
over a space-like surface of co-dimension 1
\be
\label{Q}
 Q_{\as,\at,\aA,\aB} (T|\xi )= \int_{M_{d-1}}
\tilde{J}_{\as,\at,\aA,\aB,} (T|\xi )\,,
\ee
where
$\tilde{J}_{\as,\at,\aA,\aB} (T|\xi )\,,$
is the on-mass-shell exact
$d-1$ form dual to $J^\un_{\as,\at,\aA,\aB} (T|\xi )$, i.e.
\be
J^{\un_1}_{\as,\at,\aA,\aB} (T|\xi) =\frac{1}{(d-1)!}\epsilon^{\un_1 \ldots
\un_d } \tilde{J}_{\un_2 \ldots \un_d |\,\as,\at, \aA,\aB} (T|\xi)\,.
\ee
Therefore, charges identify with the current cohomologies. The
trivial class of exact forms
$\tilde{J}_{\as,\at,\aA,\aB} (T|\xi )$ describe various ``improvements"
of the currents. In terms of the original vector currents this equivalence
has the standard form
\be
J^n \sim J^n + \partial_m H^{nm}
\ee
for any antisymmetric bivector $H^{nm}=-H^{mn}$
locally expressed in terms of the original fields. A straightforward
analysis then shows that the following equivalences take place
\be
J^{\un}_{\as,\at,\aA,\aB} (\Box T|\xi)
\sim 2(d+2\aA +2\at -4)
J^{\un}_{\as,\at,\aA-1,\aB} (T|\xi)\,,
\ee
and
\be
F^{\un }_{\as,\at,\aA,\aB} (\hat{\partial}R |\xi)
\sim - 2
G^{\un}_{\as,\at,\aA -1,\aB} (R|\xi)\,,\qquad
G^{\un }_{\as,\at,\aA,\aB} (\hat{\partial}R |\xi)
\sim - (d+2\aA+2\at -2 )
F^{\un}_{\as,\at,\aA ,\aB       } (R|\xi)\,
\ee
(with the convention that the currents with negative $\aA$
are equal to zero).

\section{AdS/CFT Correspondence}
\label{AdS/CFT Correspondence}
The parameters
$\xi^\aA_{b(\at )\,, a(\as -\aB)}$ used in (\ref{Jxi}), (\ref{Jxi1}) and
(\ref{Jxi2}) as projectors to the irreducible components of
the currents can in fact be identified with the
higher spin global symmetry parameters. This is most obvious
from the relationship between the currents and conserved charges
(\ref{Q}).
Therefore the conformal (super)currents found
in the section  \ref{Higher Spin Conformal Currents}
suggest a pattern of the global symmetry parameters of
a yet unknown conformal higher spin symmetry algebra in any dimension.
Analogously, the non-conformal (i.e. AdS or Poincare) currents
found in \cite{MV} suggest a pattern of the AdS higher spin algebra
in $d$ dimensions\footnote{Since from the
$d=4$ example it is known (see \cite{MV} and references therein) that
it is AdS geometry rather than the Poincare one that is adequate
to the higher spin dynamics at the  interaction level, the former
possibility is most interesting.}.
For the particular cases of $d=3$ or $4$ in which the
higher spin symmetries are known explicitly it was demonstrated in
\cite{MV} that indeed the one-to-one correspondence between
the patterns of the higher spin currents and AdS higher spin symmetries
takes place. Here we perform another check of the consistency of
the pattern of the higher spin currents from the perspective of the
AdS/CFT correspondence.

It is well known that the $d$-dimensional conformal algebra
$o(d,2)$ is isomorphic to the $d+1$ dimensional anti-de Sitter
algebra. The dynamical realization of this identity has been achieved
in the framework of the AdS/CFT correspondence between field theories in
the AdS space and conformal models at the boundary of the AdS space
\cite{AdS/CFT}. Let us show that the structure of the
currents obtained indicates that the analogous
correspondence is true for the higher spin symmetries.

Indeed, as shown in the section \ref{Higher Spin Conformal Currents},
conformal higher spin symmetry parameters are irreducible tensors
or tensor-spinors $\xi^\aA_{b(\at )\,, a(\as -\aB)}$
with $\as = s-1$ for integer spins $s$ and
 $\as = s-3/2$ for half-integer spins  $s$,
and arbitrary non-negative integer
parameters $\aA,\aB,\at$  satisfying the restriction
$\aA+\aB+\at\leq \as$. The integers
$\aA$, $\aB$ and $\at$ parametrize different conformal
symmetries associated with the spin $s$. On the
other hand, as shown in \cite{MV}, usual (i.e., non-conformal)
higher spin
symmetry parameters $\Xi_{m(\at ),n(\as)}$ correspond to all irreducible
two-row (tensor or tensor-spinor) Young diagrams,
for a given spin $s$ every diagram appears once ($0\leq \at\leq \as$).

Let us now assume that the parameters $\Xi_{\hm(\at ),\hn(\as)}$
describe
AdS higher spin symmetry in $d+1$ dimensions, i.e. $\hm,\hn=0\div d$,
and analyze what is a result of the dimensional reduction of this set
of parameters to $d$ dimensions. The conclusion is that the pattern
of the dimensionally reduced $AdS_{d+1}$ parameters is exactly the same
as the set of the conformal higher spin parameters
$\xi^\aA_{b(\at )\,, a(\as -\aB)}$
with allowed values of $\aA$, $\aB$ and $\at$.
To see this it is enough to observe that for the irreducible
representations of the orthogonal algebras the following
reduction rule is true
\cite{BR}
\be
\label{d+1tod}
\Xi_{\hm(\at ),\hn(\as)} \rightarrow
\sum_{\aB=0}^{\as -\at} \sum_{\aD=0}^\at \oplus
\xi_{b(\at -\aD)\,, a(\as -\aB)}
\ee
 both for (traceless) tensors and ($\gamma-$transversal) tensor-spinors.
As a result, the set of the representations of the
$d+1$-dimensional Lorentz algebra, associated
with the higher spin $AdS_{d+1}$ parameters $\Xi_{\hm(\at ),\hn(\as)}$,
exactly reproduces the set of the representations of the
$d$-dimensional Lorentz algebra associated with the
conformal higher spin parameters. This is a strong
indication that AdS/CFT correspondence extends to
the (global) higher spin symmetries.

Finally, let us note that the relationship (\ref{d+1tod}) applied
to one dimension higher tells us that the pattern of the AdS
higher spin parameters $\Xi_{m(\at ),n(\as)}$ corresponds to
the dimensional reduction of the
irreducible representation $\Xi_{M(\as),N(\as)}$
of the algebra $o(d-1,2)$ ($N,M =0 \ldots d+1$)
described by the two-row rectangular
(traceless tensor or $\gamma-$transversal tensor-spinor)
Young diagram of the length $\as$.
The same is therefore true for
the higher spin conformal symmetry parameters as resulting from
the dimensional reduction of $\Xi_{M(\as),N(\as)}$ to
two dimensions less.
This result is a natural generalization of the trivial fact that
the usual AdS and conformal symmetry parameters span the adjoint
representation of the respective orthogonal algebras,
thus corresponding to the
antisymmetric parameter $\Xi_{M,N}$, i.e. $\as=1$.

To summarize, the  higher spin extension
of $o(p,2)$ decomposes into the irreducible representations
of $o(p,2)$ described by the two-row rectangular
Young diagrams independently on whether
$o(p,2)$ is interpreted as AdS algebra (i.e., $p=d-1$) or conformal one
(i.e., $p=d$). This conclusion may be important for
the analysis of the higher spin theories in higher dimensions.

\section*{Acknowledgments}

This research was supported in part by
INTAS, Grant No.99-1-590 and by the RFBR Grants No.99-02-16207,
99-02-18414 and Grant No. 00-15-96566.


\begin{thebibliography}{77}

\bibitem{zilch} D.M.Lipkin, {\it J.Math. Phys.} {\bf 5}, 696 (1964);\\
T.A.Morgan,   {\it J.Math. Phys.} {\bf 5}, 1659 (1964);\\
D.A.Fradkin,  {\it J.Math. Phys.} {\bf 6}, 879 (1965);\\
T.W.B.Kibble, {\it J.Math. Phys.} {\bf 6}, 1022 (1965);\\
R.F. O'Connel and D.R.Tompkins, {\it J.Math. Phys.} {\bf 6}, 1952 (1965).
\bibitem{currents}
   F.A. Berends, G.J. Burgers and H. van Dam,
      {\it Nucl.~Phys.\/} B {\bf 271}, 429 (1986)\,;\\
 D. Anselmi,
      {\it Nucl.~Phys.\/} B {\bf 541}, 323 (1999)\,.
\bibitem{ex} D. Anselmi, {\it Class.Quant.Grav.} {\bf 17}, 1383 (2000)
             ({\tt hep-th/9906167}).
\bibitem{MV} M.Vasiliev,
Contributed article to Golfand's Memorial Volume
``Many faces of the superworld",
ed. by M.Shifman, World Scientific Publishing Co Pte Ltd, Singapore, 2000;
{\tt hep-th/9910096}.
\bibitem{V2}M.A. Vasiliev, {\it Fortschr. Phys.\/} {\bf 35}, 741 (1987).
\bibitem{BR} A. O.~Barut and R. R\c{a}czka,
{\it Theory of Group Representations and Applications}, PWN, Warszawa 1977.
\bibitem{AdS/CFT}
P.A.M.Dirac,  {\it J.Math.Phys.} {\bf 4}, 901 (1963);\\
J. Maldacena,
{\it Adv.Theor.Math.Phys.} {\bf 2}, 231 (1998) and
{\it Int.J.Theor.Phys.} {\bf 38}, 1113 (1998)
      ({\tt hep-th/9711200});\\
S. Ferrara and C. Fronsdal,
{\it Class.Quant.Grav.} {\bf 15}, 2153 (1998)
                  ({\tt hep-th/9712239});\\
M. Gunaydin and D. Minic,
      {\it Nucl.~Phys.\/} B {\bf 523}, 145 (1998)
      ({\tt hep-th/9802047});\\
S.S. Gubser, I.R. Klebanov, and A.M. Polyakov,
{\it Phys.Lett.} B {\bf 428}, 105 (1998)
({\tt hep-th/9802109});\\
E. Witten,
{\it Adv.Theor.Math.Phys.} {\bf 2}, 253 (1998)
      ({\tt hep-th/9802150}).


\end{thebibliography}
\end{document}